\documentstyle[12pt]{article}

\newcommand{\sect}[1]{\setcounter{equation}{0}\section{#1}}
\newcommand{\subsect}[1]{\subsection{#1}}

\newfont{\frak}{eufm10 scaled\magstep1}
\newfont{\extra}{msbm10 scaled\magstep1}

\def\be{\begin{equation}}
\def\ee{\end{equation}}
\def\bea{\begin{eqnarray}}
\def\eea{\end{eqnarray}}

\def\b{\beta}
\def\bx{\beta_x}

\def\bsx{\beta^s_x}
\def\bst{\beta^s_t}

\def\dpmx{\Delta^\pm_x}
\def\dsx{\Delta^s_x}

\def\sx{\sigma_x}
\def\st{\sigma_t}

\begin{document}

\begin{center} 
{\LARGE{\bf{Discrete derivatives \\  and \\[0.45cm] symmetries of difference
equations}}}
\end{center}

\bigskip\bigskip

\begin{center} 
D. Levi $^1$, J. Negro $^2$ and M.A. del Olmo $^2$ 
\end{center}

\begin{center} 
$^1${\sl Departimento di Fisica, Universit\'a  Roma Tre and INFN--Sezione di 
Roma Tre \\ 
Via della Vasca Navale 84, 00146 Roma, Italy}\\ 
\medskip

$^2${\sl Departamento de F\'{\i}sica Te\'orica, Universidad de Valladolid, \\
E-47011, Valladolid, Spain.}\\ 
\medskip

{e-mail:levi@amaldi.fis.uniroma3.it,  jnegro@fta.uva.es, olmo@fta.uva.es} 
\end{center}

\vskip 1.5cm
\centerline{\today}
\vskip 1.5cm

\bigskip

\begin{abstract} 
We show on the example of the discrete heat equation that for any given discrete 
derivative we can construct a nontrivial Leibniz rule 
suitable  to find the 
symmetries of  discrete equations. In this way we  obtain a symmetry Lie algebra,
defined in terms of shift operators,  isomorphic
to that of the continuous heat equation.
\end{abstract}

\vfill
\eject
\sect{Introduction\label{introduccion}}
Lie point symmetries were
introduced by Sophus Lie for solving differential equations. They turn out to 
provide one of the most efficient methods for obtaining exact analytical
solutions of partial differential equations \cite{olver}--\cite{blkumei}. This
essentially continuous method has been recently extended to the case of
discrete equations \cite{levi}--\cite{quispel}.

Let us write a general difference equation, involving, for notational 
simplicity, one scalar function  $u(x)$ of $p$ independent variables
$x=(x_1,x_2,\dots,x_p)$ evaluated at a finite number of points on a lattice.
Symbolically we write 
\be\label{ecuaciondiscreta}
E(x,\ T^{a}u(x),\ T^{b_i}\Delta_{x_i} u(x),\ T^{c_{ij}}\Delta_{x_i}\Delta_{x_j} 
u(x), \dots)=0 ,
\ee
where $E$ is some given function of its arguments,
\be
T^{a}u(x) :=\left\{ T^{a_1}_{x_1}T^{a_2}_{x_2}\cdots T^{a_p}_{x_p}\,
u(x)\right\}_{a_i=m_i}^{n_i}, \quad a=(a_1,a_2,\dots, a_p), 
\quad i=1,2,\dots, p,
\ee
with $a_i,m_i, n_i$ fixed integers ($m_i \leq n_i$),  and
\be
T^{a_i}u(x)= u(x_1,x_2,\dots,x_{i-1}, x_i+a_i\sigma_i, x_{i+1},\dots,x_p),
\ee
(the other shift operators $T^{b_i},\ T^{c_{ij}}$ are defined in a similar way),  $\Delta_{x_i}$ is
a difference operator which in the continuous limit goes into the derivative 
and $\sigma_i$ is the positive lattice  spacing in the uniform lattice of the
variable $x_i \ (i=1,\dots, p)$.  

The simplest extension of the continuous case is when the symmetry 
transformation which leaves the equation on the lattice invariant depends just
on $u(x)$ and not on its shifted values, what was called intrinsic point
transformations \cite{levi}. In this case the whole theory goes through but the
resulting transformations are somehow trivial and provide often not very
interesting solutions.

In Ref. \cite{javier} it has been proved that the intrinsic transformations can 
be extended for linear equations by considering symmetries where the transformed
function $\tilde u(x)$ depends not only on the old $u(x)$ and $x$ but also on
the function $u$ in shifted points
$T^{a_i}u(x)$ on the lattice.

The results presented in \cite{javier} where obtained considering the difference
operator
\be\label{derivadadiscretap}
\Delta_{x_i}\equiv \Delta^+_{x_i}=\frac{T_{x_i} -1}{\sigma_i}.
\ee
This difference operator is the simplest one which, when $\sigma_i \to 0$, 
goes into the standard right derivative with respect $x_i$.

Obviously, more general definitions of the difference operator can be introduced
and one would like to be able to prove that the obtained symmetries are
independent on the discretisation of the difference operator one is
considering. Among the possibilities, let us mention
\be\label{derivadadiscretam}
\Delta^-_{x_i}=\frac{1-T_{x_i}^{-1}}{\sigma_i},
\ee
corresponding to the left derivative, and the symmetric derivative 
(which goes into the derivative with respect to $x_i$ up to terms of order
$\sigma_i^2$)
\be\label{derivadadiscretas}
\Delta^s_{x_i}=\frac{T_{x_i}-T_{x_i}^{-1}}{2\sigma_i}.
\ee

Using the techniques of approximate differentiation \cite{milne} one could write
more complicate formulas for difference operators which, however, are out of the
scope of this letter.

In the following we will show that by a nontrivial definition of the Leibniz 
rule we can construct symmetries for any difference operator. Section
\ref{liesimetrias} is devoted to review the Lie group formalism for difference
equations. In Section \ref{calor} we apply it to the case of the discrete heat
equation using the discrete derivatives
(\ref{derivadadiscretam})--(\ref{derivadadiscretas}), finding difficulties in
the case (\ref{derivadadiscretas}).  In Section \ref{Leibnizcalor} we show via
examples how one can introduce in a consistent way the Leibniz rule and use
this result to obtain determining equations equivalent to those for the
standard discrete derivatives (i.e., $\Delta^\pm_{x_i}$). This fact shows that
we get different representations of the same symmetry group. In the Conclusions
we sketch how to write the determining equations when one uses a generic
difference operator  by using the  corresponding Leibniz rule obtained with
our approach.  

\sect{Lie symmetries of difference equations}\label{liesimetrias} 
Among the different algebraic methods for calculating the symmetries of discrete
equations \cite{maeda}--\cite{decio} we will make use in this paper of the
approach presented in Ref. \cite{decio}, based on the formalism of evolutionary
vector fields \cite{olver}. 

For a difference equation of order $N$ like that  given in 
(\ref{ecuaciondiscreta}) the infinitesimal symmetry vectors  in evolutionary
form, which in the continuous limit go over to point symmetries, take the
general expression
\be \label{vectorgeneral}
X_e\equiv Q\partial u= \left( \sum_i \xi _i (x, T^{a}u, \sx,\st) T^{b}
\Delta_{x_i} u -\phi(x,T^{c}u, \sx,\st)
\right)\partial u ,
\ee
with  $\xi _i (x, T^{a}u, \sx,\st)$ and $\phi(x,T^{c}u, \sx,\st)$ functions 
which in the continuous limit go over to $\xi _i (x, u)$ and $\phi(x, u)$,
respectively.

The vector fields $X_e$ generate the symmetry group of the discrete equation 
(\ref{ecuaciondiscreta}), whose elements transform solutions $u(x)$ of the
equation  into solutions $\tilde u(x)$.
The $N$--th prolongation of $X_e$ must verify the invariance condition
\be\label{prolongacion}
pr^N X_e E|_{E=0}=0.
\ee
The group generated by the prolongations also transforms solutions into 
solutions, and  $\Delta_{x_i}u,\Delta_{x_i}\Delta_{x_j}u, \dots $ (up to order
$N$) into the variations of  $\tilde u$ with respect to $x_i$. The formula of
$pr^N X_e$ is given by
\be\label{prolongationvector}
pr^N X_e=\sum_a T^a Q\partial_{T^a u} +
\sum_{b_i} T^{b_i}Q^{x_i}\partial_{T^{b_i}\Delta_{x_i}u}  +
\sum_{c_{ij}} T^{c_{ij}}Q^{x_{i}x_{j}}\partial_{T^{c_{ij}}
\Delta_{x_i}\Delta_{x_j}u}  +
\dots
\ee
The summations in (\ref{prolongationvector}) are over all the sites present in
(\ref{ecuaciondiscreta}), $Q^{x_i},\, Q^{{x_i}{x_j}},\dots$ are total 
variations of $Q$, i.e.,
\be
 Q^{x_i}=\Delta_{x_i}^T Q, \qquad 
 Q^{{x_i}{x_j}}=\Delta_{x_i}^T \Delta_{x_j}^T Q ,\qquad \cdots,
\ee
where the partial variation $\Delta_{x_i}$ is defined by
\be\begin{array}{ll}
\Delta_{x} f(x, u(x), \Delta_{x}u(x), \dots)=&\frac 1\sigma[f(x+\sigma, u(x),
(\Delta_{x}u)(x), \dots) \\[0.2cm] 
&\qquad  -f(x, u(x), \Delta_{x}u(x), \dots)], \qquad x=x_i ,
\end{array}\ee
and the total variation $\Delta_{x_i}^T$ by
\be\begin{array}{ll}
\Delta_{x}^T f(x, u(x), \Delta_{x}u(x), \dots)=&\frac 1\sigma[f(x+\sigma, 
u(x+\sigma),
(\Delta_{x}u)(x+\sigma), \dots) \\[0.2cm] 
&\qquad -f(x, u(x), \Delta_{x}u(x), \dots)], \qquad \forall x=x_i .
\end{array}\ee
 
The symmetries of the equation (\ref{ecuaciondiscreta}) are given by equation
(\ref{prolongacion}). From it we get the determining equations for $\xi _i$ 
and $\phi$  as the coefficients of linearly independent expressions in the
discrete derivatives $T^{a}\Delta_{x_i}u$, $T^{b}\Delta_{x_ix_j}u$, \dots.

The Lie commutators of the vector fields $X_e$ are obtained by commuting their 
first prolongations and projecting onto the symmetry algebra $\cal G$, i.e.,
\be\begin{array}{lll}\label{comnmutador}
[X_{e1},X_{e2}] &=& [pr^{1}X_{e1},pr^{1}X_{e2}]|_{\cal G}\\ [0.2cm]
&=&\left(
Q_1\frac{\partial Q_2}{\partial u}  -Q_2\frac{\partial Q_1}{\partial u} 
+ Q^{x_i}_1\frac{\partial Q_2}{\partial u_{x_i}} 
-Q_2^{x_i}\frac{\partial Q_1}{\partial u_{x_i}}  \right )\partial_u \ ,
\end{array}\ee
where the $\partial u_{x_i}$ terms  disappear after projection onto $\cal G$. 

The formalism presented above is very complicate and if the system is 
nonlinear is almost impossible to get a result since the number of terms to
consider is a priori infinite. The situation is simpler in the case  of linear
equations for which we can use a reduced Ansatz.
We assume that the  evolutionary vectors
(\ref{vectorgeneral}) have the form 
\be\label{vectorrestringido}
X_e= \left( \sum_i \xi _i (x, T^{a}, \sx,\st) 
\Delta_{x_i} u -\phi(x,T^{a}, \sx,\st) u \right)\partial_u .
\ee
Now the vector fields $X_e$ can be written as
$X_e= (\hat X u) \partial_u$,  i.e., 
\be\label{vectorlie}
\hat X=  \sum_i \xi _i (x, T^{a}, \sx,\st) 
\Delta_{x_i}  -\phi(x,T^{a}, \sx,\st)  .
\ee
The operators $\hat X$ may span a subalgebra of the Lie symmetry algebra 
(see Ref. \cite{decio}).

\sect{Discrete heat equation}\label{calor}
Let us consider the equation
\be\label{calorecuacion0}
(\Delta_{t}-\Delta_{xx})u(x)=0 ,
\ee
which is a discretisation of the heat equation. As the equation is linear we 
can consider an evolutionary vector field of the form
\be\label{evolvector}
X_e\equiv Q\partial_u=(\tau\Delta_{t} +\xi \Delta_{x} u+ f u) \partial_u ,
\ee
with $\tau$, $\xi$ and $f$ arbitrary functions of $x, t, T_x$ $T_t$, $\sx$ 
and $\st$. As $T_x$ and $T_t$ are operators not commuting  with $x$ and $t$,
respectively, $\tau$, $\xi$ and $f$ are operator valued functions. Since
equation (\ref{calorecuacion0}) is a second order difference equation it is
necessary to use the second prolongation. The determining equation is
\be\label{ecuacioncalordet0}
\Delta_{t}^T Q -  \Delta^T_{xx} Q|_{\Delta_{xx}u=\Delta_{t} u}=0 ,
\ee
which explicitly  reads
\be\label{ecuacioncalordet1}
\Delta_{t}(\xi\Delta_{x}u) +\Delta_{t}(\tau\Delta_{t}u)+ \Delta_{t}(fu)
-[\Delta_{xx}(\xi\Delta_{x}u) +\Delta_{xx}(\tau\Delta_{t}u)+ \Delta_{xx}(fu)] 
|_{\Delta_{xx}u=\Delta_{t} u}=0 .
\ee

We had no need to give the explicit form of the
operator $\Delta$ to get equation (\ref{ecuacioncalordet1}). Only when 
developing expression (\ref{ecuacioncalordet1}) we need to apply Leibniz rule
and, hence, the results will depend from the given definition of the discrete
derivative.

\subsect{Symmetries in the right (left) discrete derivative case}
\label{simetriasdiscetasderecha}

Choosing as in Ref. \cite{javier,decio} the derivative $\Delta^+$ and 
consequently the Leibniz rule 
\be\label{Leibnizrulep}
\Delta^+ (fg)=\Delta^+ (f) Tg +f\Delta^+ g 
\ee 
we obtain the following set of determining equations
\be\begin{array}{l}\label{ecuacacionesdeterminantescalor}
\Delta_{x}^+\tau=0 ,\\[0.2cm]
(\Delta_{t}^+\tau)T_t -2 (\Delta^+_{x}\xi) T_x =0 ,\\[0.2cm]
(\Delta_{t}^+\xi) T_t - (\Delta^+_{xx}\xi) T_x^2  -2 (\Delta^+_{x}f) T_x =0 ,
\\[0.2cm]
(\Delta^+_{t} f) T_t -(\Delta^+_{xx}f)T_x^2=0 ,
\end{array}\ee 
by equating to zero  the coefficients of $\Delta _{xt} u$, 
$\Delta _{t} u$,  $\Delta _{x} u$ and $u$, respectively.
The solution of  (\ref{ecuacacionesdeterminantescalor}) gives
\be\label{calorsolucionesdeterminingequation}
\begin{array}{l}
\tau=t^{(2)}\tau_2 + t \tau_1 +\tau_0 ,\\[0.2cm]
\xi=\frac 12 x(\tau_1+2t\tau_2)T_tT_x^{-1}+t\xi_1+\xi_0, \\[0.2cm]
f=\frac 14 x^{(2)} \tau_2T_t^2T_x^{-2}+\frac 12 t\tau_2T_t+
\frac 12 x\xi_1T_tT_x^{-1}+\gamma ,
\end{array}\ee 
where $\tau_0,\  \tau_1,\  \tau_2,\  \xi_0,\ \xi_1$ and $\gamma$ are 
arbitrary functions of $T_x$, $T_t$ and of the spacings $\sigma_x$ and
$\sigma_t$, and $x^{(n)}$, $t^{(n)}$ are the Pochhammer symbols given by, for
instance, 
\be
x^{(n)}=x(x-\sigma_x)\dots (x-(n-1)\sigma_x).
\ee 

By a suitable choice of the functions  $\tau_i,\ \xi_i,$ and $\gamma$ we get 
the following symmetries
\be\label{caloralgebra}
\begin{array}{ll}
P_0= (\Delta_t u)\partial_u ,& \qquad (\tau_0 =1)\\[0.2cm]
P_1=(\Delta_x u)\partial_u , & \qquad (\xi_0 =1)\\[0.2cm]
W= u\partial_u , & \qquad (\gamma =1)\\[0.2cm]
B=(2tT_t^{-1}\Delta_x u +xT_x^{-1} u )\partial_u , & \qquad (\xi_1 =2 T_t^{-1})
\\[0.2cm] 
D= (2tT_t^{-1}\Delta_t u +xT_x^{-1}\Delta_x u + \frac 12 u)\partial_u , &
\qquad (\tau_1 =2 T_t^{-1},\gamma =\frac12)\\[0.2cm]  
K=(t^2T_t^{-2}\Delta_t u - \sigma_t t T_t^{-2}\Delta_t u +
txT_t^{-1}T_x^{-1}\Delta_x u & \\[0.2cm] 
\qquad \qquad +  \frac 14 x^2 T_x^{-2} u  - \frac14 \sigma_x x T_x^{-2}u
+\frac12 tT_t^{-1} u)\partial_u  ,&
\qquad (\tau_2 = T_t^{-2})
\end{array}\ee 
that close into a 6--dimensional Lie algebra which is isomorphic to the 
symmetry algebra of the continuous heat equation.
\bigskip

A second choice for the discrete derivative is $\Delta^-$ and now the Leibniz 
rule becomes
\be\label{Leibnizrulem}
\Delta^- (fg)=\Delta^- (f) T^{-1}g +f\Delta^- g
\ee 
This gives the same results (\ref{calorsolucionesdeterminingequation}) and
(\ref{caloralgebra})  provided 
we make the substitution
$T \rightarrow T^{-1}$.

\subsect{Symmetries in the symmetric discrete derivative case}
\label{simetriasdiscetassimetrica}

Next let us consider the case of the symmetric derivative $\Delta^s$. It seems
natural to propose the following  Leibniz rule 
 \be\label{Leibnizrules}
\Delta^s (fg)=\Delta^s (f) T^{-1}g +(Tf)\Delta^s g .
\ee 
In this case taking as before coefficients of the different discrete 
derivatives of $u$  ($\Delta _{tt} u$, $\Delta _{xt} u$, $\Delta _{t}u$, 
$\Delta_{x} u$  and $u$, respectively) the  determining equations are
\be\begin{array}{l}\label{ecuacacionesdeterminantescalor3}
T_t\tau -T_x^2\tau=0 ,\\[0.2cm]
T_t\xi -T_x^2\xi -2 (T_x\Delta^s_{x}\tau)T_x^{-1}=0 ,\\[0.2cm]
(\Delta^s_{t}\tau)T_t^{-1}+(T_tf) -2 (T_x\Delta^s_{x}\xi) T_x ^{-1}
-(\Delta^s_{xx}\tau)
T_x^{-2}- T_x^2 f=0 ,\\[0.2cm] 
(\Delta^s_{t}\xi) T_t ^{-1} - (\Delta^s_{xx}\xi) T_x^ {-2} -2
(T_x\Delta^s_{x}f) T_x ^{-1} =0 ,\\[0.2cm] 
(\Delta^s_{t} f) T_t ^{-1} -(\Delta^s_{xx}f)T_x^ {-2}=0 .
\end{array}\ee 
Note that there is one equation (the first one, associated to $\Delta _{tt} u$) 
more that in the two previous cases. This implies that the solution of such
equations (\ref{ecuacacionesdeterminantescalor3}) is just
\be
\tau=\tau_0 ,\quad
\xi=\xi_0, \quad f=f_0 , 
\ee 
with $\tau_0,\  \xi_0,$ and $f_0$ are arbitrary functions of $T_x$,
$T_t$ and of the spacings $\sigma_x$ and $\sigma_t$.  

Obviously, in this last case our naive approach does not allows us to recover 
the whole symmetry algebra of the heat equation (\ref{caloralgebra}).  In fact,
we will show in next Section that the decomposition  of equation
(\ref{ecuacioncalordet1}) into equations (\ref{ecuacacionesdeterminantescalor3})
via   Leibniz rule (\ref{Leibnizrules}) is not the most appropriate one.

\sect{Leibniz rule and symmetries with symmetric \\derivatives}
\label{Leibnizcalor}

Let us consider a formal way to get Leibniz rule
in the continuous case which is easily extendible to the case of  discrete 
derivatives. This result will allow us to get the symmetries of a discrete
equation independently of the discrete derivative used. 

Starting from the well known result $[\partial_x,x]=1$ by algebraic methods we get
that  $[\partial_x,f(x)]=f'(x)$ (at least for analytic   functions). Consequently,
the Leibniz rule  $\partial_x (f g)=fg'+f' g$ can be derived. 

Now, let us consider the commutator with the  differential operator substituted 
by a difference one. By direct computation we have
\be\label{relacionesconmutaciondiscretas0}
[\Delta^\pm_x,x]=T^{\pm 1}_x ,\qquad
[\Delta^s_x,x]=\frac{T_x+T_x^{-1}}{2}.
\ee
By introducing a function $\bx=\b(T_x)$ we can always rewrite the commutation 
relations  (\ref{relacionesconmutaciondiscretas0}) as
\be\label{relacionesconmutaciondiscretas1}
[\Delta_x,x\bx]=1 .
\ee
Let us note that $\Delta_x\bx=\bx\Delta_x$ and $\Delta_t\bx=\bx\Delta_t$.

For the standard left and right derivatives $\Delta^{\pm 1}_x$ we easily find 
that  $\bx^{\pm}=T^{\mp 1}_x$, and  we get that
\be
[\Delta^{\pm}_x,f(x)\b^{\pm}]=\left(\Delta^{\pm}_xf(x)\right).
\ee
From this last expression we recover  Leibniz rules (\ref{Leibnizrulep}) and
(\ref{Leibnizrulem}).  

For the  symmetric discrete derivative $\Delta^{s}_x$ we obtain the formal expression
\be
\bsx=2(T_x+T_x^{-1})^{-1},
\ee 
and 
\be\label{commutadorfs}
[\Delta^s_x,f(x)\bsx]=\left(\Delta^s_xf(x)\right)T_x\bsx
+\left(T_x^{-1} f(x)-f(x)\right)\bsx \Delta^s_x.
\ee
From  relation (\ref{commutadorfs}) we get the Leibniz rule
\be\begin{array}{c}\label{Leibnizsimetrica2}
\Delta^s_x\left(f(x)g(x)\right)= f(x) \Delta^s_x g(x)+\left[\right.
\frac{1}{\sigma_x}\left((T_x^{-1}-1)f(x)\right)
\left(T_x-(\bsx)^{-1}\right)\\[0.2cm]
 \qquad \qquad +\left. \left(\Delta^s_xf(x)\right)T_x\right]g(x).
\end{array}\ee
Let us note that expression (\ref{Leibnizsimetrica2}) looks very different from 
(\ref{Leibnizrules}).

Formula (\ref{Leibnizsimetrica2}) allows us to  write down the 
determining equations
(\ref{ecuacioncalordet1}) as 
\be\begin{array}{l}\label{ecuacacionesdeterminantescalor4}
\left((1-T_x^{-1})\tau\right) T_x^{-1} +\left((T_x-1)\tau\right) T_x=0 ,
\\[0.3cm]
\frac{1}{2\st}\left[\left((1-T_t^{-1})\tau\right)T_t^{-1} +
\left((T_t-1)\tau\right)T_t \right] \\ [0.2cm]\qquad\qquad
- \frac{1}{\sx}\left[\left((1-T_x^{-1})\xi\right)T_x^{-1}
+\left((T_x-1)\xi\right)T_x\right]=0,\\[0.3cm]
\frac{1}{2\st}\left[\left((1-T_t^{-1})\xi\right)T_t^{-1}
+\left((T_t -1)\xi\right)T_t \right] -
\frac{1}{\sx}\left[\left((1-T_x^{-1})f\right)T_x^{-1}+
\left((T_x -1)f\right)T_x 
 \right]\\ [0.2cm]\qquad\qquad-
\frac{1}{4\sx^2} \left[\left((1-T_x^{-1})^{2}\xi\right)T_x^{-2}
+2\left((T_x+T_x^{-1}-2)\xi\right) +\left((T_x-1)^{2}\xi\right)T_x^{2}\right]=0,
\\[0.3cm]
\frac{1}{2\st}\left[ \left((1-T_t^{-1})f\right)T_t^{-1}
+\left((T_t -1)f\right)T_t \right]\\[0.2cm]
\qquad\qquad 
-\frac{1}{4\sx^2} \left[\left((1-T_x^{-1})^{2}f\right)T_x^{-2}
+2\left((T_x+T_x^{-1}-2)f\right) +\left((T_x-1)^{2}f\right)T_x^{2}\right] = 0 ,
\end{array}\ee 
obtained as coefficients of $\Delta _{xt} u$, $\Delta _{t} u$, $\Delta
_{x} u$  and $u$, respectively. So, we get the same number of equations as 
in the cases of the standard discrete derivatives $\Delta^\pm$.

The solution of eqs.(\ref{ecuacacionesdeterminantescalor4}) is given by
\be\label{ecuacacionesdeterminantescalor4soluciones}\begin{array}{l}
\tau^s = t^{(2)}\tau_2 +t\tau_1+\tau_0 ,
\\[0.2cm]
\xi^s = \frac {1}{2}x \left(2t\tau_2 + \tau_1 + \st T_t^{-1} \bst \tau_2\right)
(\bst)^{-1}\bsx +t\xi_1+\xi_0,
\\[0.2cm]
f^s = \frac{1}{4} x^{(2)} \tau_2(\bsx)^2 (\bst)^{-2}  
{+} \frac{1}{2}x\xi_1 \bsx (\bst)^{-1} 
{+} \frac 14 x\sx\tau_2 T_x^{-1} (\bsx)^3(\bst)^{-2} {+} \frac12 t \tau_2
 (\bst)^{-1}  {+} f_0,
\end{array}\ee 
where $\tau_2$, $\tau_1$, $\tau_0$, $\xi_1$, $\xi_0$ and $f_0$ are arbitrary 
functions of $T_x,\ T_t,\ \sx$ and $\st$.

From  (\ref{ecuacacionesdeterminantescalor4soluciones}) and 
(\ref{evolvector})  we obtain, with a suitable choice of 
$\tau_2$, $\tau_1$, $\tau_0$, $\xi_1$, $\xi_0$ and $f_0$, the following
symmetries
\be\label{caloralgebrasimetrica}
\begin{array}{ll}
P_0^s= (\Delta^s_t u)\partial_u ,&\quad  (\tau_0=1)\\[0.2cm]
P_1^s=(\Delta^s_x u)\partial_u ,&\quad  (\xi_0=1) \\[0.2cm]
W^s= u\partial_u , &\quad  (f_0=1)\\[0.2cm]
B^s=(2t\bst\Delta^s_x u + x\bsx u )\partial_u ,  
&\quad  (\xi_1=2\bst)\\[0.2cm]
D^s= (2t\bst\Delta^s_t u +x\bsx\Delta^s_x u + \frac 12 u)\partial_u ,
&\quad  (\tau_1=2\bst, f_0 =\frac12)\\[0.2cm]
K^s=((t^2(\bst)^{2}- t{\sigma_t}^2(\bst)^3{\Delta^s_t})\Delta^s_t u  +
tx\bst\bsx\Delta^s_x u   &\quad  (\tau_2={\bst}^2,\\[0.2cm]
\qquad \quad  
- \frac 14 x{\sigma_x}^2(\bsx)^3\Delta_x^s u + \frac 14 x^2 (\bsx)^{2} u
+ \frac 12 t\bst u)\partial_u , 
&\quad\ \  \tau_1=\sigma_t{\bst}^2{-}\sigma_t^2{\bst}^3\Delta^s_t).
\end{array}\ee 
They close the same  6--dimensional Lie algebra generated by the operators 
(\ref{caloralgebra}). The above result deserves some comments. First, there
appear functions $\bst$ ($\bsx$) of $T_t$ ($T_x$) that can only be understood
as infinite series developments. Therefore, some of the above symmetries
(\ref{caloralgebrasimetrica}) have not a local character in the sense that they
are not polynomials in  $T_t^{\pm1},T_x^{\pm1}$. Although such symmetries give
rise to the classical symmetries in the limit $\sigma_x\to 0,\sigma_t\to 0$,
one of them ($K$) also includes surprisingly a term in
$(\Delta_t)^2$ (which vanishes in the continuous limit since it has a factor
$\sigma_t^2$).

\sect{Conclusions}\label{conclusiones}

Analysing the different Leibniz rules used in Sections
\ref{calor} and \ref{Leibnizcalor}, we see that when we get the correct
result the Leibniz rule must have the form
\be\label{Leibnizgeneral}
\Delta_x\left(f(x)g(x)\right)= f(x) \Delta_x g(x)+
D_x(f(x))g(x),
\ee
where $D_x$ is a function of $T_x,\ \bx$ and $\sx$ (we should have  
written $D_x(f(x);T_x,\bx,\sx)$, with $\beta_{x}$ given by 
eq.\ (\ref{relacionesconmutaciondiscretas1}), but for the sake
of shortness we will simply write $D_x(f)$, and similarly for $D_t(f)$). Once
we have chosen a  particular discrete derivative $\Delta_x$ we can
write the explicit expression of $D_x$. 
In particular, for $\dpmx$ and $\dsx$ their corresponding $D_x(f)$ functions are
\be\begin{array}{lll}\label{Leibnizgeneralx}
D_x^\pm(f)&=&(\Delta_x^\pm (f)) (\bx^\pm)^{-1}
=(\Delta_x^\pm (f)) T_x^{\pm 1}, \\[0.2cm] 
D_x^s(f)&=&\frac{1}{\sigma_x}
\left((T_x^{-1}-1)f\right)\left(T_x-(\bsx)^{-1}\right)
 + \left(\Delta^s_xf\right)T_x\\[0.2cm] 
&=&\left.\left( (T_x^{-1}-1)f\right)\Delta_x^s+ \left((\Delta_x^s)f \right)
T_x\right]
\\[0.2cm] 
&=&\frac{1}{2\sigma_x}\left[\left( (1-T_x^{-1})f\right)T_x^{-1}
+ \left((T_x-1)f \right)T_x\right] .
\end{array}\ee 

Using the general Leibniz rule (\ref{Leibnizgeneral}) for an arbitrary discrete
derivative we obtain from (\ref{ecuacioncalordet1}), equating to zero  the
coefficients of $\Delta _{xt} u$, $\Delta _{t} u$,  $\Delta _{x} u$ and $u$,
respectively, the following set of  determining equations
\be\begin{array}{l}\label{ecuacacionesdeterminantescalorgeneral}
D_{x}(\tau)=0 ,\\[0.2cm]
D_{t}(\tau) -2 D_{x}(\xi)=0,\\[0.2cm]
D_{t}(\xi) - D_{xx}(\xi)  -2 D_{x}(f)=0 ,\\[0.2cm]
D_{t} (f)  -D_{xx}(f)=0 .
\end{array}\ee 
where $D_{xx}(f)=D_{x}(D_{x}(f))$.

For the cases $\Delta^\pm$ and
$\Delta^s$  from  (\ref{ecuacacionesdeterminantescalorgeneral}) we recover the
determining equations  (\ref{ecuacacionesdeterminantescalor}) and
(\ref{ecuacacionesdeterminantescalor4}), respectively. 

We have always four equations, whose general solution will depend  on the formal
expression of the corresponding  $\Delta$. In fact, as long as we are able to find
for the discrete derivative $\Delta$ the operator $\b$ satisfying the commutator
(\ref{relacionesconmutaciondiscretas1}) this will allow us to give particular
solutions to the determining equations (\ref{ecuacacionesdeterminantescalorgeneral})
keeping the same Lie structure of the symmetries of the classical heat equation. Of
course, some of these discrete symmetries can have a rather complicated expression
with a nonlocal character, but this feature is a consequence of the structure of the
operator $\b$.

This procedure can be straightforwardly applied to other discretisations such as
the wave equation \cite{nn96} or even equations including a potential term. 

Something similar happens with  discrete equations on a
$q$-lattice: once we have a solution for $\b$ we can derive symmetry operators with
a Lie structure (a situation which is seldom considered). On the other hand, if we
are more interested in local symmetries, with some restrictions, the
natural structure is that of a $q$--algebra. 
However, these issues are out of our present scope and they will be published
elsewhere.

\section*{Acknowledgments}
This work has been partially supported by 
DGES of the  Ministerio de Educaci\'on y Cultura of Spain under 
Projects PB98-0360 and the Junta de Castilla y Le\'on (Spain).  The 
visit of DL to Valladolid and of MAO to Rome have been partially financed by 
the Erasmus European Project.


\end{document}